\documentclass{article}
\def\Ref#1{(\ref{#1})}
\usepackage{amsmath}
\usepackage{cite}
\begin{document}
\begin{titlepage}
\noindent{\large \bf Exactly solvable models through the empty
interval method, for more-than-two-site interactions}

\vskip 2 cm

\begin{center}{M. Khorrami$^{a,d}${\footnote
{mamwad@iasbs.ac.ir}} , A. Aghamohammadi$^{b,d}${\footnote
{mohamadi@azzahra.ac.ir}}\& M. Alimohammadi$^{c,d}${\footnote
{alimohmd@ut.ac.ir}} } \vskip 5 mm

{\it{ $^a$ Institute for Advanced Studies in Basic Sciences,
             P.~O.~Box 159,\\ Zanjan 45195, Iran. }

{ $^b$ Department of Physics, Alzahra University,
             Tehran 19834, Iran. }

{ $^c$ Physics Department, University of Tehran,
             North Karegar Avenue, Tehran, Iran. }

{ $^d$ Institute of Applied Physics, IUST,
             P.~O.~Box 16845-163, Tehran, Iran. }
             }
\end{center}

\begin{abstract}
\noindent Single-species reaction-diffusion systems on a
one-dimensional lattice are considered, in them more than two
neighboring sites interact. Constraints on the interaction rates
are obtained, that guarantee the closedness of the time evolution
equation for $E_n(t)$'s, the probability that $n$ consecutive
sites are empty at time $t$. The general method of solving the
time evolution equation is discussed. As an example, a system with
next-nearest-neighbor interaction is studied.
\end{abstract}
\begin{center} {\bf PACS numbers:} 05.40.-a, 02.50.Ga

\end{center}

\end{titlepage}
\newpage
\section{Introduction}
In contrast to equilibrium systems, which are best analyzed using
standard equilibrium statistical mechanics, there is no general
approach to study the systems far from equilibrium. People are
motivated to study the non-equilibrium systems in one dimension,
since these are in principle easier. Different methods have been
used to study stochastic models in one dimension, including
analytical and asymptotic methods, mean-field methods, and
large-scale numerical methods. Some models solved using these
methods, are studied for example in
\cite{ScR,ADHR,KPWH,HS1,PCG,HOS1,HOS2,AL,RK,AKK2,MAM}.

There is no universal meaning for the term exactly solvable. For
example in \cite{GS,AAMS,SAK}, solvability means that the
evolution equation of $n$-point functions contains only $n$- or
less- point functions. In \cite{AA,RK3}, solvability means that
the $S$-matrix of the $N$-particle system is factorized into
products of 2-particle $S$-matrices. This means that the
$S$-matrices should satisfy the Yang-Baxter equation. Another
meaning of integrability is that the time evolution equation for
$E_n(t)$, the probability that $n$ consecutive sites are empty at
time $t$, is closed, that is it can be expressed in terms of other
$E_m(t)$'s. This method of solving the integrable models is called
the empty interval method (EIM).

The empty interval method has been used to analyze the one
dimensional dynamics of diffusion-limited coalescence
\cite{BDb,BDb1,BDb2,BDb3}. Using this method, the functions
$E_n(t)$ have been calculated.  For the cases of finite
reaction-rates, some approximate solutions have been obtained. EIM
has been also generalized to study the kinetics of the $q$-state
one-dimensional Potts model in the zero-temperature limit
\cite{Mb}.

In \cite{AKA},  all one dimensional reaction-diffusion models with
nearest-neighbor interactions, exactly-solvable through EIM, have
been studied. In \cite{HH}, EIM has also been used to study a
specific model with next-nearest-neighbor interaction. In
\cite{MB}, the conventional EIM has been extended to a more
generalized form. Using this extended version, a model not
solvable by conventional EIM has been studied.

In this article, we consider systems, in them more than two
neighboring sites interact. We consider the most general systems
with $k$-site interactions. Some constraints are imposed on the
interaction rates, so that the time evolution equation for
$E_n(t)$ is closed. The general method of solving the time
evolution equation is also discussed. Finally, as an example, a
system with next-nearest-neighbor interactions has been considered
in more detail.

\section{Models solvable through the empty interval method}
Consider a general one-species reaction-diffusion model on a
one-dimensional periodic lattice with $L+1$ sites, with a
$k$-neighboring-site interaction. We want to find criteria on the
interaction rates, that guarantee the solvability of the system
via EIM, that is, the closedness of the evolution equation for the
probability that $n$ consecutive sites are empty, $E_n$.

Suppose that the initial condition of the system is
translationally-invariant. Any configuration of $k$ neighboring
sites is denoted by $\mathbf{ a}=( a_1, a_2,\cdots, a_k)$, where $
a_i=\circ$ or $\bullet$. $\circ (\bullet )$ is used to denote an
empty (occupied) site. The rate of transition from a configuration
$\mathbf{ a}$ to $\mathbf{ b}$ is denoted by $\lambda_{\mathbf{
a}}^{\mathbf{ b}}$. Similar to \cite{AKA}, the interactions with $k$ empty
sites as initial or final configuration are not considered here. In other
words for any $\mathbf{a}$,
\begin{equation}\label{1}
\lambda_{\mathbf{0}}^{\mathbf{ a}}= \lambda_{\mathbf{
a}}^{\mathbf{0}}=0.
\end{equation}
Excluding these interactions from the $2^k(2^k-1)$ possible
interactions, $(2^k-1)(2^k-2)$ interactions remain to be
considered. We want to impose restrictions on $\lambda_{\mathbf{
a}}^{\mathbf{ b}}$'s in such a way that the evolution equation for
$E_n(t)$'s be closed. As we will see, the form of evolution
equation generally will be different for $n\geq k-1$ and $n<k-1$,
and also will be different for $n+k>L+2$ and $n+k\leq L+2$. So we
will treat each case separately.
\subsection{The case $n\geq k-1$ and $n+k\leq L+2$}
To obtain evolution equation for $E_n(t)$, one should first
recognize the source and sink terms. There are two cases. In the first case,
the intersection of the empty block and the interacting block is
in the left-hand side of the empty block. In the other case, this
intersection is in the right-hand side of the empty block.
For the first case, the source terms come from
\begin{equation}\label{2}
  { a'}_1\cdots{ a'}_l\overbrace{ c_1\cdots
   c_{k-l}\circ\cdots\circ}^n\to  b_1\cdots
   b_l\overbrace{\circ\cdots\circ}^n,
\end{equation}
where $\mathbf{ c}\ne \mathbf{0}$. Here $\mathbf{0}$ stands for a
block of adjacent empty sites. One also has $l\leq k-1$.
$\lambda^{\mathbf{0}}_{\mathbf{ a}}=0$ leads to $l\geq 1$.
So the left source for $E_n$ is
\begin{equation}\label{3}
  S_{L}=\sum_{l=1}^{k-1}
  \sum_{{\mathbf{a'},\mathbf{b}}\atop{\mathbf{c}\ne\mathbf{0}}}
  \lambda_{\mathbf{ a'}\mathbf{ c}}^
  {\mathbf{ b}\mathbf{0}}
  P(\mathbf{ a'}\mathbf{ c}\overbrace{\circ\cdots\circ}^{n-k+l}).
\end{equation}
Now consider the expansion
\begin{align}\label{4}
 \sum_{{\mathbf{a'}}\atop{\mathbf{c}\ne
  \mathbf{0}}}\lambda^{\mathbf{f}}_{\mathbf{ a'},\mathbf{ c}}
  =&\sum_{\mathbf{a}_1}\lambda^{\mathbf{f}}_{\mathbf{ a}_1\bullet}
  +\sum_{\mathbf{ a}_2}\lambda^{\mathbf{f}}_{\mathbf{ a}_2\bullet\circ}
  +\cdots +
  \sum_{\mathbf{a'}}\lambda^{\mathbf{f}}_{\mathbf{a'}\bullet\mathbf{0}}
  \nonumber\\
  =&\sum_{l'=l}^{k-1}
  \sum_{\mathbf{ a}}
  \lambda^{\mathbf{f}}_{\mathbf{ a}\bullet\mathbf{0}},
\end{align}
where in the last equality, $\mathbf{a}$ is an $l'$-dimensional
vector and $\mathbf{0}$ is $(k-l'-1)$-dimensional. So,
\begin{equation}\label{5}
  S_L=\sum_{l=1}^{k-1}\sum_{l'=l}^{k-1}
  \sum_{\mathbf{ a},\mathbf{ b}}
  \lambda_{\mathbf{ a}\bullet \mathbf{0}}
  ^{\mathbf{ b}\mathbf{0}}
  P(\mathbf{ a}\bullet\overbrace{\circ\cdots\circ}^{n+l-l'-1}).
\end{equation}
If
\begin{equation}\label{6}
  \Lambda^L_{ll'}:=\sum_{\mathbf{b}}
  \lambda_{ a_1\cdots a_{l'}\bullet\mathbf{0}}^{
   b_1\cdots b_{l}\mathbf{0}},\qquad 1\leq l\leq
  k-1,\quad l\leq l'\leq k-1,
\end{equation}
is independent of $\mathbf{ a}$, then one can sum up $P(\mathbf{
a}\bullet \overbrace{\circ\cdots\circ}^n)$ on the index $\mathbf{a}$. Then
\begin{align}\label{7}
  S_L&=\sum_{l=1}^{k-1}\sum_{l'=l}^{k-1}\Lambda^L_{ll'}P(\bullet
  \overbrace{\circ\cdots\circ}^{n+l-l'-1})\nonumber\\
  &=\sum_{l=1}^{k-1}\sum_{l'=l}^{k-1}\Lambda^L_{ll'}(E_{n+l-l'-1}-E_{n+l-l'}).
\end{align}

One can do similar calculations for the case that the intersection of the
interaction block and the empty block is in the right-hand side of the
empty block. Defining
\begin{equation}\label{8}
  \Lambda^R_{ll'}:=\sum_{\mathbf{b}}
  \lambda_{\mathbf{0}\bullet a_1\cdots a_{l'}}^{
  \mathbf{0} b_1\cdots b_{l}},\qquad 1\leq l\leq
  k-1,\quad l\leq l'\leq k-1,
\end{equation}
and assuming that it is independent of  $\mathbf{ a}$, the
source term for this case is
\begin{equation}\label{9}
  S_R=\sum_{l=1}^{k-1}\sum_{l'=l}^{k-1}\Lambda^R_{ll'}P(
\overbrace{\circ\cdots\circ}^{n+l-l'-1}\bullet).
\end{equation}
Putting these together, the source term is
\begin{equation}\label{10}
  S=\sum_{l=1}^{k-1}\sum_{l'=l}^{k-1}(\Lambda^L_{ll'}+\Lambda^R_{ll'})
  (E_{n+l-l'-1}-E_{n+l-l'}).
\end{equation}

Now, lets consider the sink terms. Again we will treat
interactions of left- and right-hand sides separately. First
consider the left ones. The interactions which contribute to sink
terms come from
\begin{equation}\label{11}
{ a'}_1\cdots{ a'}_l\overbrace{\circ\cdots\circ}^n\to  b_1\cdots
 b_l c_1\cdots  c_{k-l}\overbrace{\circ\cdots\circ}^{n-k+l},
\end{equation}
where $\mathbf{c}\ne\mathbf{0}$. The sink term from the left
interactions is
\begin{equation}\label{12}
  R_L=-\sum_{l=1}^{k-1}
  \sum_{{\mathbf{a'},\mathbf{b}}\atop
  \mathbf{c}\ne\mathbf{0}}
  \lambda_{\mathbf{a'}\mathbf{0}}
  ^{\mathbf{b}\mathbf{c}}
  P(\mathbf{a'}\overbrace{\circ\cdots\circ}^n).
\end{equation}
$\lambda_{\mathbf{0}}^{\mathbf{a}}=0$ leads to $l\geq 1$. One also
has $l\leq k-1$. $\lambda_{\mathbf{ a}}^{\mathbf{ b}}$ is the
transition rate, so it is defined only for ${\mathbf{
a}}\ne{\mathbf{ b}}$. But one can extend this definition and define the
diagonal terms in such a way that
\begin{equation}\label{13}
\sum_{\mathbf{b},\mathbf{c}}\lambda_{\mathbf{a}\mathbf{d}}^{
\mathbf{b}\mathbf{c}}=0.
\end{equation}
Then it is seen that
\begin{equation}\label{14}
\sum_{{\mathbf{b}}\atop{\mathbf{c}\ne\mathbf{0}}} \lambda_{\mathbf{
a}\mathbf{ d}}^{ \mathbf{ b}\mathbf{ c}}=- \sum_{\mathbf{ b}}
\lambda_{\mathbf{ a}\mathbf{ d}}^{ \mathbf{ b}\mathbf{0}}.
\end{equation}
Using this, one arrives at the following equation for $R_L$
\begin{equation}\label{15}
  R_L=\sum_{l=1}^{k-1}
  \sum_{\mathbf{a'},\mathbf{b}}
  \lambda_{\mathbf{a'}\mathbf{0}}
  ^{\mathbf{b}\mathbf{0}}
  P(\mathbf{a'}\overbrace{\circ\cdots\circ}^n).
\end{equation}
This again recasts to a simpler form, using the following
expansion
\begin{align}\label{16}
 \sum_{\mathbf{ a'}}
 \lambda^{\mathbf{f}}_{{ a'}_1\cdots { a'}_l \mathbf{0}}
  =&\sum_{\mathbf{a'}}\lambda^{\mathbf{f}}_{
  a'_1\cdots a'_{l-1}\bullet\mathbf{0}}
  +\sum_{\mathbf{a'}}\lambda^{\mathbf{f}}_{
  a'_1\cdots a'_{l-2}\bullet\mathbf{0}}+\cdots +
  \lambda^{\mathbf{f}}_{\bullet\mathbf{0}}\nonumber\\
   =&\sum_{l'=0}^{l-1}
  \sum_{\mathbf{ a}}
  \lambda^{\mathbf{f}}_{a_1\cdots a_{l'}\bullet\mathbf{0}}.
\end{align}
Putting these together, one arrives at
\begin{equation}\label{17}
  R_L=\sum_{l=1}^{k-1}\sum_{l'=0}^{l-1}\Lambda^L_{ll'}P(\bullet
  \overbrace{\circ\cdots\circ}^{n+l-l'-1}),
\end{equation}
where
\begin{equation}\label{18}
  \Lambda^L_{ll'}:=\sum_{\mathbf{b}}
  \lambda_{ a_1\cdots a_{l'}\bullet\mathbf{0}}^{
   b_1\cdots b_{l}\mathbf{0}},\qquad 1\leq l\leq k-1,\quad 0\leq l'\leq l-1,
\end{equation}
and it is assumed that $\Lambda^L_{ll'}$ is independent of
$\mathbf{ a}$.

It is seen that the conditions we have obtained for the source and
sink terms for the left interactions, eqs. \Ref{6} and \Ref{18},
are similar, except for the range of $l'$. Performing similar
calculations for the right interactions, all the conditions coming
from the source and sink terms can be summarized as this. The
following quantities should be independent of $\mathbf{a}$.
\begin{align}\label{19}
  \Lambda^L_{ll'}:=&\sum_{\mathbf{b}}
  \lambda_{ a_1\cdots a_{l'}\bullet\mathbf{0}}^{
   b_1\cdots b_{l}\mathbf{0}},\qquad 1\leq l\leq
  k-1,\quad 0\leq l'\leq k-1,\\
\Lambda^R_{ll'}:=&\sum_{\mathbf{b}}
  \lambda_{\mathbf{0}\bullet a_1\cdots a_{l'}}^{
  \mathbf{0} b_1\cdots b_{l}},\qquad 1\leq l\leq
  k-1,\quad 0\leq l'\leq k-1.
\end{align}
Defining $\Lambda_{ll'}:=\Lambda^L_{ll'}+\Lambda^R_{ll'}$, the
time evolution equation of  $E_n(t)$, for $n\geq k-1$ and $n+k\leq
L+2$, takes the following form
\begin{equation}\label{20}
  {{\rm d}E_n(t)\over {\rm
  d}t}=\sum_{l=1}^{k-1}\sum_{l'=0}^{k-1}
  \Lambda_{ll'}(E_{n+l-l'-1}-E_{n+l-l'}).
\end{equation}
Note that in this equation, $E_0$ is defined through
\begin{equation}
E_0:=1.
\end{equation}
\subsection{The case $n<k-1$ and $n+k\leq L+2$}
Now, we want to derive time evolution equation of $E_n(t)$'s when
$n<k-1$. Two cases may occur. The first one, is that the $n$
adjacent  sites which we are focused on, are among the $k$
interacting sites, and in the second case a block of these sites
is outside those $k$ sites. The result for the second case is
similar to that of the preceding subsection, $n\geq k-1$. For that
case, we only quote the results. However, we study the first case
in more detail.

The source terms come from
\begin{equation}\label{21}
  { a_1}'\cdots{ a_p}' c_1\dots c_n{e_1}'
  \cdots{e_q}'\to  b_1\cdots b_p\overbrace{\circ\circ\cdots
  \circ}^n d_1\cdots d_q,
\end{equation}
where $\mathbf{ c}\ne\mathbf{0}$, $p+q+n=k$, and $p,q\geq 1$.
Then the source term is
\begin{equation}\label{22}
 S=\sum_{{p,q=1}\atop{p+q=k-n}}\sum_{{\mathbf{a'},\mathbf{e'},\mathbf{b}
 ,\mathbf{d}}\atop{\mathbf{c}\ne\mathbf{0}}}\lambda_{\mathbf{ a'}\mathbf{ c}
\mathbf{e'}}^{\mathbf{ b}\mathbf{0} \mathbf{d}}P(\mathbf{a'ce'}).
\end{equation}
Similar to the previous cases, one can rearrange the sum of the
rates in the following form.
\begin{equation}\label{23}
  \sum_{{\mathbf{a'},\mathbf{e'}}\atop{\mathbf{c}\ne
  \mathbf{0}}}\lambda_{\mathbf{a'}\mathbf{c}\mathbf{e'}}^{\mathbf{f}}=
  \sum_{n'=0}^{n-1}\sum_{n''=0}^{q-1}\sum_{\mathbf{a},\mathbf{e}}
  \lambda_{\mathbf{a}\bullet\underbrace{\circ\cdots\circ}_{n'+n''}\bullet
  \mathbf{e}}^{\mathbf{f}}+\sum_{n'=0}^{n-1}\sum_{\mathbf{a}}
  \lambda_{\mathbf{a}\bullet\mathbf{0}}^{\mathbf{f}},
\end{equation}
where we have used the fact that $\mathbf{ c}\ne\mathbf{0}$ and so
at least one of the $ c_i$'s should be $\bullet$. In the above
equation, $\mathbf{ a}$ is a $(p+n-n'-1)$-dimensional vector and
$\mathbf{e}$ is a $(q-n''-1)$-dimensional vector. In the first
term on the right-hand side of the above equation, the left
$\bullet$ is the the first $\bullet$ in $\mathbf{c}$ from the
right, and the right $\bullet$ is the first $\bullet$ in
$\mathbf{e'}$ from the left. In the second term, the $\bullet$ is
the the first $\bullet$ in $\mathbf{c}$ from the right, and
$\mathbf{e'}$ is $\mathbf{0}$. Arranging all these together, one
arrives at the following equation for the source term.
\begin{equation}\label{24}
S=\sum_{{p,q=1}\atop{p+q=k-n}}\sum_{n'=0}^{n-1}\left[
\sum_{n''=0}^{q-1}\sum_{\mathbf{a},\mathbf{e},\mathbf{b},
\mathbf{d}}
  \lambda_{\mathbf{a}\bullet\mathbf{0}\bullet
\mathbf{e}}^{\mathbf{b}\mathbf{0}\mathbf{d}}
P(\mathbf{a}\bullet\mathbf{0}\bullet\mathbf{e})+
\sum_{\mathbf{a},\mathbf{b},\mathbf{d}}
  \lambda_{\mathbf{ a}\bullet\mathbf{0}}^{\mathbf{
   b}\mathbf{0}
\mathbf{ d}}P(\mathbf{a}\bullet\mathbf{0})\right].
\end{equation}
Defining
\begin{equation}\label{25}
  \Lambda_{pq,p'q'}:=\sum_{\mathbf{b},\mathbf{d}}
  \lambda_{ a_1\cdots a_{p'}\bullet\mathbf{0}\bullet
  e_1\cdots e_{q'}}^{ b_1\cdots b_p\mathbf{0} d_1\cdots d_q},
\end{equation}
and
\begin{equation}\label{26}
  \Lambda^L_{pq,p'}:=\sum_{\mathbf{b},\mathbf{d}}
  \lambda_{ a_1\cdots a_{p'}\bullet\mathbf{0}}^{ b_1\cdots b_p
  \mathbf{0} d_1\cdots d_q},
\end{equation}
where $p\leq p'\leq p+n-1$, $0\leq q'\leq q-1$, $p+q=k-n$, and
$p,q\geq 1$. Assuming that $\Lambda_{pq,p'q'}$ is independent of
$\mathbf{a}$ and $\mathbf{e}$ and $\Lambda^L_{pq,p'}$ is
independent of $\mathbf{a}$, one can sum up the terms in
(\ref{24}):
\begin{equation}\label{27}
S=\sum_{{p,q=1}\atop{p+q=k-n}}\sum_{n'=0}^{n-1}\left[
\sum_{n''=0}^{q-1}\Lambda_{pq,p'q'}
P(\bullet\underbrace{\circ\cdots\circ}_{n'+n''}\bullet )+
\Lambda^L_{pq,p'}P(\bullet\underbrace{\circ\cdots\circ}_{q+n'})
\right],
\end{equation}
or in terms of $E_n$'s,
\begin{align}\label{28}
S=&\sum_{{p,q=1}\atop{p+q=k-n}}\sum_{n'=0}^{n-1}\Bigg[
\sum_{n''=0}^{q-1}\Lambda_{pq,p'q'}\left(E_{n'+n''}+
E_{n'+n''+2}-2E_{n'+n''+1}\right)
\nonumber \\
&+\Lambda^L_{pq,p'}\left(E_{q+n'}-E_{q+n'+1}\right)\Bigg].
\end{align}
The independency of $\Lambda_{pq,p'q'}$ with respect to
$\mathbf{a}$ and $\mathbf{e}$, and $\Lambda^L_{pq,p'}$ with
respect to $\mathbf{a}$ is sufficient to guarantee that the above
source term is expressible in terms of $E_n$'s, but is not
necessary. For example, in (\ref{22}) one can decompose the blocks
$\mathbf{c}$ and $\mathbf{a'}$ instead of $\mathbf{c}$ and
$\mathbf{e'}$, which leads to another set of sufficient conditions
on the rates.

Now, lets consider the sink terms for $n<k-1$:
\begin{equation}\label{29}
   a'_1\cdots a'_p\overbrace{\circ\cdots\circ}^n e'_1\cdots e'_q
  \rightarrow b_1\cdots b_p c_1\cdots c_n d_1\cdots d_q.
\end{equation}
The above interaction produces a sink term:
\begin{equation}\label{30}
R=-\sum_{{p,q=1}\atop{p+q=k-n}}\sum_{{\mathbf{a'},\mathbf{e'},
 \mathbf{b},\mathbf{d}}\atop{\mathbf{c}\ne\mathbf{0}}}
 \lambda_{\mathbf{ a'}\mathbf{0}
\mathbf{e'}}^{\mathbf{ b}\mathbf{ c} \mathbf{ d}}P(\mathbf{
a'}\mathbf{0}\mathbf{e'}).
\end{equation}
Similar to the preceding case, using (\ref{14}), one arrives at
\begin{equation}\label{31}
R=\sum_{{p,q=1}\atop{p+q=k-n}}\sum_{\mathbf{a'},\mathbf{e'},
 \mathbf{b},\mathbf{d}}\lambda_{\mathbf{ a'}\mathbf{0}\mathbf{e'}}^
 {\mathbf{ b}\mathbf{0} \mathbf{ d}}P(\mathbf{
a'}\mathbf{0}\mathbf{e'}).
\end{equation}
Using the expansion
\begin{align}\label{32}
\sum_{\mathbf{ a'},\mathbf{e'}}\lambda_{\mathbf{ a'}
\mathbf{0}\mathbf{e'}}^{\mathbf{f}}=&\sum_{\mathbf{ a}
,\mathbf{e}}\sum_{q'=0}^{q-1}\sum_{p'=0}^{p-1}\lambda_{a_1\cdots
a_{p'} \bullet\mathbf{0}\bullet e_1\cdots e_{q'}}^{\mathbf{f}}\nonumber\\
&+\sum_{\mathbf{e}}\sum_{q'=0}^{q-1}\lambda_{ \mathbf{0}\bullet
e_1\cdots e_{q'}}^{\mathbf{f}} + \sum_{\mathbf{ a}
}\sum_{p'=0}^{p-1}\lambda_{a_1\cdots a_{p'} \bullet
\mathbf{0}}^{\mathbf{f}},
\end{align}
$R$ can be written in the form
\begin{align}\label{33}
  R=\sum_{{p,q=1}\atop{p+q=k-n}}\Bigg[&\sum_{q'=0}^{q-1}
  \sum_{p'=0}^{p-1}\Lambda_{pq,p'q'}
  (E_{k-p'-q'-2}+E_{k-p'-q'}-2E_{k-p'-q'-1})\nonumber\\
  &+\sum_{q'=0}^{q-1}\Lambda^R_{pq,q'}
  (E_{k-q'-1}-E_{k-q'})\nonumber\\
  &+\sum_{p'=0}^{p-1}\Lambda^L_{pq,p'}
  (E_{k-p'-1}-E_{k-p'})\Bigg],
\end{align}
where we have used the definition (\ref{25}) and (\ref{26}) for
$\Lambda_{pq,p'q'}$ and $\Lambda^L_{pq,p'}$ but with an extension
of the range of $p'$ and $q'$ to $0\leq p'\leq p+n-1$ and $0\leq
q'\leq q-1$. It has been also assumed that $\Lambda_{pq,p'q'}$ is
independent of $\mathbf{a}$ and $\mathbf{e}$, and
$\Lambda^L_{pq,p'}$ is independent of $\mathbf{e}$.
$\Lambda^R_{pq,q'}$ is defined through
\begin{equation}\label{34}
\Lambda^R_{pq,q'}:=\sum_{\mathbf{b},\mathbf{d}}\lambda_{\mathbf{0}\bullet
e_1\cdots e_{q'}}^{b_1\cdots b_p\mathbf{0}d_1\cdots d_q},\qquad
p+q=k-n,\quad 0\leq q'\leq q-1,
\end{equation}
and it is assumed that it is independent of $\mathbf{e}$.
Considering (\ref{28}), (\ref{33}), and the source- and sink-terms
corresponding to the previous subsection, and noting that in these
latter terms, one should replace $1\leq l\leq k-1$ in the
right-hand side of (\ref{20}) with $k-n\leq l\leq k-1$, one
arrives (for $n<k-1$ and $n+k\leq L+2$) at
\begin{align}\label{34b}
  {{\rm d}E_n(t)\over {\rm
  d}t}=&\sum_{l=k-n}^{k-1}\sum_{l'=0}^{k-1}
  \Lambda_{ll'}(E_{n+l-l'-1}-E_{n+l-l'})\nonumber\\
  &+\sum_{{p,q=1}\atop{p+q=k-n}}
  \Bigg\{\sum_{n'=0}^{p-1}\Bigg[\sum_{n''=0}^{q-1}
  \Lambda_{pq,(p+n-n'-1)(q-n''-1)}\nonumber\\
  &\times (E_{n'+n''}+E_{n'+n''+2}-2E_{n'+n''+1})\nonumber\\
  &+\Lambda^L_{pq,p+n-n'-1}(E_{q+n'}-E_{q+n'+1})\Bigg]\nonumber\\
  &+\sum_{n'=0}^{p-1}\sum_{n''=0}^{q-1}\Lambda_{pq,n'n''}
  (E_{k-n'-n''-2}+E_{k-n'-n''}-2E_{k-n'-n''-1})\nonumber\\
  &+\sum_{n'=0}^{p-1}\Lambda^L_{pq,n'}
  (E_{k-n'-1}-E_{k-n'})\nonumber\\
  &+\sum_{n''=0}^{q-1}\Lambda^R_{pq,n''}
  (E_{k-n''-1}-E_{k-n''})\Bigg\}.
\end{align}
\subsection{The case $n+k>L+2$}
The last case to be considered is the case with $n+k>L+2$.
Normally the case of large $L$ and finite $k$ is of interest, in
which one also has $n>k$. We assumed periodic boundary condition
for the system. Then the intersection of the $k$ interacting sites
and the block of $n$ sites may consist of two disconnected parts,
of lengths $l$ and $l'$. So, one has, in addition to the source
terms similar to those of subsection 2.1, a source term coming
from
\begin{equation}\label{35}
  \underbrace{a'_1\cdots a'_l\circ\cdots\circ b'_1\cdots b'_{l'}}_n
  {c'}_1\cdots c'_{k-l-l'}\to \mathbf{0}d_1\cdots d_{k-l-l'}.
\end{equation}
This leads to a source term
\begin{equation}\label{37}
  S=\sum_{{l,l'=1}\atop{l+l'=n+k-L-1}}
  \sum_{{\mathbf{a'},\mathbf{b'},\mathbf{c'},\mathbf{d}}\atop
   {\mathbf{a'}\ne\mathbf{0}\ \mathrm{or}\ \mathbf{b'}\ne\mathbf{0}}}
   \lambda_{\mathbf{b'}\mathbf{c'}\mathbf{a'}}^{\mathbf{0}
   \mathbf{d}\mathbf{0}}P(\mathbf{a'}
   \overbrace{\circ\cdots\circ}^{n-l-l'}\mathbf{b'}\mathbf{c'}).
\end{equation}
Using
\begin{align}\label{38}
  \sum_{{\mathbf{a'}\mathbf{b'}\mathbf{c'}}\atop{\mathbf{a'}\ne
  \mathbf{0}\ \mathrm{or}\ \mathbf{b'}\ne\mathbf{0}}}\lambda
  ^{\mathbf{f}}_{b'_1\cdots b'_{l'}\mathbf{c'}a'_1\cdots
  a'_l}=&\sum_{p=0}^{l-1}\sum_{\mathbf{c}}
\lambda_{\underbrace{\circ\cdots\circ}_{l'}\mathbf{c}\bullet
\underbrace{\circ\cdots\circ}_p}^{\mathbf{f}}\nonumber\\
&+\sum_{q=0}^{l'-1}\sum_{\mathbf{c}}
\lambda_{\underbrace{\circ\cdots\circ}_q\bullet\mathbf{c}
\underbrace{\circ\cdots\circ}_l}^{\mathbf{f}}\nonumber\\
&+\sum_{p=0}^{l-1}\sum_{q=0}^{l'-1}\sum_{\mathbf{c}}
\lambda_{\underbrace{\circ\cdots\circ}_q\bullet\mathbf{c}\bullet
\underbrace{\circ\cdots\circ}_p}^{\mathbf{f}},
\end{align}
it is seen that if the quantities
\begin{align}\label{40}
\Lambda'_{ll',pq}:=
&\sum_{\mathbf{d}}\lambda_{\underbrace{\circ\cdots\circ}_q\bullet
\mathbf{c}\bullet\underbrace{\circ\cdots\circ}_p}^{
\overbrace{\circ\cdots\circ}^{l'}\mathbf{d}
\overbrace{\circ\cdots\circ}^{l}},\quad 0\leq p\leq l-1,
\quad 0\leq q\leq l'-1\nonumber\\
\Lambda^{\prime L}_{ll',p}:=&\sum_{\mathbf{d}}
\lambda_{\underbrace{\circ\cdots\circ}_{l'}\mathbf{c}\bullet
\underbrace{\circ\cdots\circ}_p}^{\overbrace{\circ\cdots\circ}^{l'}
\mathbf{d}\overbrace{\circ\cdots\circ}^l},
\quad 0\leq p\leq l-1\nonumber\\
\Lambda^{\prime R}_{ll',q}:=
&\sum_{\mathbf{d}}\lambda_{\underbrace{\circ\cdots\circ}_q\bullet
\mathbf{c}\underbrace{\circ\cdots\circ}_l}
^{\overbrace{\circ\cdots\circ}^{l'}
\mathbf{d}\overbrace{\circ\cdots\circ}^l},\quad 0\leq q\leq l'-1
\end{align}
are independent of $\mathbf{c}$, then the source term
corresponding to (\ref{35}) is
\begin{align}\label{440}
S=&\sum_{{l,l'=1}\atop{l+l'=n+k-L-1}}\Bigg\{
\sum_{p=0}^{l-1}\sum_{q=0}^{l'-1}\Lambda'_{ll',pq}(-2
E_{n+p+q-l-l'+1}+E_{n+p+q-l-l'}+E_{n+p+q-l-l'+2})\nonumber\\
&+\sum_{p=0}^{l-1}\Lambda^{\prime L}_{ll',p}
(E_{n+p-l}-E_{n+p-l+1})\nonumber\\
&+\sum_{q=0}^{l'-1}\Lambda^{\prime R}_{ll',q}
(E_{n+q-l'}-E_{n+q-l'+1})\Bigg\}.
\end{align}

Now let's consider the sink terms. Again there are terms similar
to of subsection 2.1, and a new sink term, which is
\begin{equation}\label{42}
  R=-\sum_{{l,l'=1}\atop{l+l'=n+k-L-1}}
  \sum_{{\mathbf{a'},\mathbf{b},\mathbf{c},\mathbf{d}}\atop
  {\mathbf{b}\ne\mathbf{0}\ \mathrm{or}\
  \mathbf{d}\ne\mathbf{0}}}\lambda_{\underbrace{\circ\cdots\circ}_{l'}
  \mathbf{a'}\underbrace{\circ\cdots\circ}_{l}}^{b_1\cdots
  b_{l'}\mathbf{c}d_1\cdots
  d_l}P(\mathbf{a'}\overbrace{\circ\cdots\circ}^n).
\end{equation}
Using
\begin{equation}\label{43}
  \sum_{{\mathbf{b},\mathbf{c},\mathbf{d}}
  \atop{\mathbf{b}\ne\mathbf{0}\ \mathrm{or}\
  \mathbf{d}\ne\mathbf{0}}}\lambda_{\mathbf{a}}^
  {\mathbf{b},\mathbf{c},\mathbf{d}}=-
   \sum_{\mathbf{c}}\lambda_{\mathbf{a}}^{\mathbf{0}
   \mathbf{c}\mathbf{0}}
\end{equation}
$R$ can be written as
\begin{equation}\label{44}
R=\sum_{{l,l'=1}\atop{l+l'=n+k-L-1}} \sum_{\mathbf{a'},\mathbf{c}}
\lambda_{\underbrace{\circ\cdots\circ}_{l'}\mathbf{a'}
\underbrace{\circ\cdots\circ}_l}^
{\overbrace{\circ\cdots\circ}^{l'}\mathbf{c}
\overbrace{\circ\cdots\circ}^l}
P(\mathbf{a'}\overbrace{\circ\cdots\circ}^n).
\end{equation}
Since $\mathbf{a'}\ne \mathbf{0}$ (eq. (\ref{1})), one has
expansion
\begin{equation}\label{45}
\sum_{\mathbf{a'}}\lambda^{\mathbf{f}}
_{\underbrace{\circ\cdots\circ}_{l'}
\mathbf{a'}\underbrace{\circ\cdots\circ}_l}=\sum_{q=0}^{k-l-l'}
\sum_{\mathbf{a}}\lambda^{\mathbf{f}}_
{\underbrace{\circ\cdots\circ}_{l'}a_1\cdots a_q\bullet
\mathbf{0}}.
\end{equation}
If
\begin{equation}
\Lambda^{\prime L}_{ll',p}:=\sum_{\mathbf{c}}
\lambda_{\underbrace{\circ\cdots\circ}_{l'}\mathbf{a}\bullet
\underbrace{\circ\cdots\circ}_p}
^{\overbrace{\circ\cdots\circ}^{l'}
\mathbf{c}\overbrace{\circ\cdots\circ}^l}, \quad l\leq p\leq
k-l'-1
\end{equation}
is independent of $\mathbf{a}$, then the above sink term becomes
\begin{equation}\label{46}
  R=\sum_{{l,l'=1}\atop{l+l'=n+k-L-1}}
  \sum_{p=l}^{k-l'-1}\Lambda^{\prime L}_{ll',p}
  (E_{n+p-l}-E_{n+p-l+1}).
\end{equation}
Note that here too, this condition on $\Lambda^{\prime L}$ is a
sufficient condition for the EIM-solvability of the model.
Using (\ref{440}), (\ref{46}), and the source- and sink-terms
corresponding to those of subsection 2.1, one arrives at
\begin{align}\label{48}
{{{\rm d}E_n}\over{{\rm d}t}}=&\sum_{l=1}^{L-n-1}\sum_{l'=0}^{k-1}
\Lambda_{ll'}(E_{n+l-l'-1}-E_{n+l-l'})\nonumber\\
&+\sum_{{l,l'=1}\atop{l+l'=n+k-L-1}}\Bigg[\sum_{p=0}^{l-1}
\sum_{q=0}^{l'-1}\Lambda'_{ll',pq}(E_{L-k+p+q+1}+
E_{L-k+p+q+3}-2 E_{L-k+p+q+2})\nonumber\\
&+\sum_{p=0}^{k-l'-1}\Lambda^{\prime L}_{ll',p}(E_{n+p-l}-
E_{n+p+1-l})\nonumber\\
&+\sum_{q=0}^{l'-1}\Lambda^{\prime R}_{ll',q}(E_{n+q-l'}-
E_{n+q+1-l'})\Bigg]
\end{align}
for $n+k>L+2$ (and $n>k$). Note that the summation limits in the
terms corresponding to the source and sink terms coming from the
processes investigated in subsection 2.1, have been properly
modified.
\section{General method of the solution}
In the previous section, the evolution equation of $E_n$'s were
obtained, eqs. (\ref{20}), (\ref{34b}), and (\ref{48}).
Investigating (\ref{34b}) and (\ref{48}), one can see that these
equations can be rewritten in the general form of (\ref{20}),
provided one defines $E_n$'s for $n<0$, and $n>L+1$ properly.
Doing this, one arrives at
\begin{equation}\label{48b}
  {{\rm d}E_n(t)\over {\rm
  d}t}=\sum_{l=1}^{k-1}\sum_{l'=0}^{k-1}
  \Lambda_{ll'}(E_{n+l-l'-1}-E_{n+l-l'}),
\end{equation}
for any $n$, with the following constraints (which are actually
definitions).
\begin{equation}\label{48c}
\sum_{s=r}^{k-1} M_{rs}(E_s-E_{s+1})=0,\qquad -k+2\leq r\leq -1,
\end{equation}
and
\begin{equation}\label{48d}
\sum_{s=L+2-k}^{r} N_{rs}(E_{s-1}-E_s)=0,\qquad L+2\leq r\leq
L+k-1.
\end{equation}
In addition to these, there are two other boundary conditions
\begin{equation}\label{480}
E_0=1,
\end{equation}
and
\begin{equation}\label{481}
E_{L+1}=0.
\end{equation}
This last condition comes from the fact that if the lattice is
initially nonempty, it will never become empty (as it is seen from
(\ref{1})). So, excluding the empty lattice (which remains empty)
there will always be at least one particle on the lattice.
Equations (\ref{48c}) to (\ref{481}) are $2k-2$ boundary condition
for the difference equation (\ref{48b}), which is of the same
order $2k-2$. To solve these equations, first consider the
stationary solution. This solution ($E^{\mathrm P}_n$) satisfies
\begin{equation}\label{482}
  \sum_{l=1}^{k-1}\sum_{l'=0}^{k-1}\Lambda_{ll'}
  (E^{\mathrm P}_{n+l-l'-1}-E^{\mathrm P}_{n+l-l'})=0,
\end{equation}
with the same boundary conditions (\ref{48c}) to (\ref{481}). The
solution to (\ref{482}) is
\begin{equation}\label{483}
 E^{\mathrm P}_n=\sum_{p=1}^{2k-2}\alpha_p z_p^n,
\end{equation}
where $z_p$'s are the solutions of
\begin{equation}\label{484}
  \sum_{l=1}^{k-1}\sum_{l'=0}^{k-1}\Lambda_{ll'}
  (z^{l-l'-1}-z^{l-l'})=0.
\end{equation}
This equation has $2k-2$ roots, one of them is $1$. The
coefficients $\alpha_p$ can be determined using the constraints
(\ref{48c}) to (\ref{481}):
\begin{align}\label{485}
\sum_{s=r}^{k-1}\sum_{p=1}^{2k-2}M_{rs}\alpha_p
(z_p^s-z_p^{s+1})&=0,\qquad -k+2\leq r\leq -1,\nonumber\\
\sum_{s=L+2-k}^{r}\sum_{p=1}^{2k-2}N_{rs}\alpha_p
(z_p^{s-1}-z_p^s)&=0,\qquad L+2\leq r\leq L+k-1,\nonumber\\
\sum_{p=1}^{2k-2}\alpha_p&=1,\nonumber\\
\sum_{p=1}^{2k-2}\alpha_p z_p^{L+1}&=0.
\end{align}

The full solution is of the form
\begin{equation}\label{486}
E_n(t)=:E_n^{\mathrm P}+F_n(t),
\end{equation}
where $F_n(t)$ satisfies an equation similar to (\ref{48b}) but
with homogeneous boundary conditions:
\begin{align}\label{487}
  {{\rm d}F_n(t)\over {\rm
  d}t}&=\sum_{l=1}^{k-1}\sum_{l'=0}^{k-1}
  \Lambda_{ll'}(F_{n+l-l'-1}-F_{n+l-l'}),\nonumber\\
\sum_{s=r}^{k-1} M_{rs}(F_s-F_{s+1})&=0,\qquad -k+2\leq r\leq
-1,\nonumber\\
\sum_{s=L+2-k}^{r} N_{rs}(F_{s-1}-F_s)&=0,\qquad L+2\leq r\leq
L+k-1,\nonumber\\
F_0&=0,\nonumber\\
F_{L+1}&=0.
\end{align}
To solve this, one writes $F_n$ as
\begin{equation}\label{488}
F_n(t)=\sum_{\epsilon}e^{\epsilon t}F_{\epsilon,n},
\end{equation}
where $F_{\epsilon,n}$ satisfies
\begin{align}\label{489}
  \epsilon F_{\epsilon,n}&=\sum_{l=1}^{k-1}\sum_{l'=0}^{k-1}
  \Lambda_{ll'}(F_{\epsilon,n+l-l'-1}-F_{\epsilon,n+l-l'}),\nonumber\\
\sum_{s=r}^{k-1} M_{rs}(F_{\epsilon,s}-F_{\epsilon,s+1})&=0,\qquad
-k+2\leq r\leq
-1,\nonumber\\
\sum_{s=L+2-k}^{r}
N_{rs}(F_{\epsilon,s-1}-F_{\epsilon,s})&=0,\qquad L+2\leq r\leq
L+k-1,\nonumber\\
F_{\epsilon,0}&=0,\nonumber\\
F_{\epsilon,L+1}&=0.
\end{align}
$F_{\epsilon,n}$ can be written as
\begin{equation}\label{490}
  F_{\epsilon,n}=\sum_{p=1}^{2k-2}\beta_{\epsilon,p}z_{\epsilon,p}^n,
\end{equation}
where $z_{\epsilon,p}$'s should satisfy
\begin{equation}\label{491}
  \sum_{l=1}^{k-1}\sum_{l'=0}^{k-1}\Lambda_{ll'}
  (z^{l-l'-1}-z^{l-l'})=\epsilon.
\end{equation}
This equation has $2k-2$ roots. The coefficients
$\beta_{\epsilon,p}$ satisfy
\begin{align}\label{492}
\sum_{s=r}^{k-1}\sum_{p=1}^{2k-2}M_{rs}\beta_{\epsilon,p}
(z_{\epsilon,p}^s-z_{\epsilon,p}^{s+1})&=0, \qquad -k+2\leq r\leq -1,
\nonumber\\
\sum_{s=L+2-k}^{r}\sum_{p=1}^{2k-2}N_{rs}\beta_{\epsilon,p}
(z_{\epsilon,p}^{s-1}-z_{\epsilon,p}^s)&=0,\qquad L+2\leq r\leq
L+k-1,
\nonumber\\
\sum_{p=1}^{2k-2}\beta_{\epsilon,p}&=0,\nonumber\\
\sum_{p=1}^{2k-2}\beta_{\epsilon,p}z_{\epsilon,p}^{L+1}&=0.
\end{align}
These are a set of $2k-2$ linear homogeneous equations for the
$2k-2$ variables $\beta_{\epsilon,p}$. The condition that there
exists a nonzero solution for these variables is that the
determinant of matrix of coefficients be zero. This is a condition
for $\epsilon$. So, in principle, one can solve this equation to
obtain the solutions for $\epsilon$, and then the corresponding
solution for $z_{\epsilon,p}$'s. One can then obtain
$\beta_{\epsilon,p}$'s, and $F_n(t)$ is obtained using (\ref{490})
and (\ref{488}).
\section{A model with three-site interaction}
As an example, consider a model with three sites
(next-nearest-neighbor) interaction. Denoting the eight possible
three-state configurations as following
\begin{align}\label{50}
\mathbf{0}:=(\circ\circ\circ)\qquad
\mathbf{1}:=(\circ\circ\bullet)\qquad
\mathbf{2}:=(\circ\bullet\circ)\qquad
\mathbf{3}:=(\circ\bullet\bullet)\nonumber\\
\mathbf{4}:=(\bullet\circ\circ)\qquad
\mathbf{5}:=(\bullet\circ\bullet)\qquad
\mathbf{6}:=(\bullet\bullet\circ)\qquad
\mathbf{7}:=(\bullet\bullet\bullet)\qquad,
\end{align}
and the transition-rate from the state $i$ to the state $j$ by
$\lambda_i^j$, one can write the conditions for the solvability of
the system through the empty-interval method as
\begin{align}\label{51}
&\lambda_6^4=\lambda_2^4,\nonumber\\
&\lambda_3^1=\lambda_2^1,\nonumber\\
&\lambda_7^4=\lambda_5^4=\lambda_3^4=\lambda_1^4,\nonumber\\
&\lambda_7^1=\lambda_6^1=\lambda_5^1=\lambda_4^1,\nonumber\\
&\lambda_2^1+\lambda_2^3+\lambda_2^5+\lambda_2^7=
\lambda_6^1+\lambda_6^3+\lambda_6^5+\lambda_6^7,\nonumber\\
&\lambda_3^7+\lambda_3^6+\lambda_3^5+\lambda_3^4=
\lambda_2^7+\lambda_2^6+\lambda_2^5+\lambda_2^4,\nonumber\\
&\lambda_1^2+\lambda_1^6=\lambda_3^2+\lambda_3^6=
\lambda_5^2+\lambda_5^6=\lambda_7^2+\lambda_7^6,\nonumber\\
&\lambda_7^3+\lambda_7^2=\lambda_6^3+\lambda_6^2=
\lambda_5^3+\lambda_5^2=\lambda_4^3+\lambda_4^2,\nonumber\\
&\lambda_7^5+\lambda_7^1=\lambda_3^5+\lambda_3^1,\nonumber\\
&\lambda_6^5+\lambda_6^1=\lambda_2^5+\lambda_2^1,\nonumber\\
&\lambda_7^2=\lambda_5^2,\nonumber\\
&\lambda_3^2=\lambda_1^2,\nonumber\\
&\lambda_6^2=\lambda_4^2.
\end{align}
For example, independence of $\Lambda_{11}^L$ with respect to
$\mathbf{a}$, gives $\lambda_2^4=\lambda_6^4$. One, of course, has
also
\begin{equation}\label{100}
\lambda_0^i=\lambda_i^0=0.
\end{equation}
This is nothing but eq. (\ref{1}). Using (\ref{20}) for
$1=k-2<n<L-k+3=L$, we have
\begin{align}\label{49}
  {{\rm d}E_n(t)\over {\rm  d}t}=&\sum_{l=1}^{2}\sum_{l'=0}^{2}
  \Lambda_{ll'}(E_{n+l-l'-1}-E_{n+l-l'})\nonumber\\
  =&-\Lambda_{20}E_{n+2}+(-\Lambda_{10}+\Lambda_{20}-\Lambda_{21})
  E_{n+1}+(\Lambda_{10}-\Lambda_{11}+\Lambda_{21}-\Lambda_{22})E_{n}
  \nonumber\\
  &+(\Lambda_{11}-\Lambda_{12}+\Lambda_{22})E_{n-1}+
  \Lambda_{12}E_{n-2},\qquad 1<n<L.
\end{align}
The time-evolution equations for $E_1$ and $E_{L}$ come from
(\ref{34b}) and (\ref{48}), respectively:
\begin{align}\label{49b}
{{\rm d}E_1(t)\over {\rm
d}t}=&\sum_{l'=0}^2\Lambda_{2l'}(E_{2-l'}-E_{3-l'})\nonumber\\
      &+\Lambda_{11,10}(E_0+E_2-2 E_1)+\Lambda^L_{11,1}(E_1-E_2)\nonumber\\
      &+\Lambda_{11,00}(E_1+E_3-2 E_2)+\Lambda^L_{11,0}(E_2-E_3)\nonumber\\
      &+\Lambda^R_{11,0}(E_2-E_3),
\end{align}
and
\begin{align}\label{49c}
{{\rm d}E_{L}(t)\over {\rm
d}t}=&\sum_{l'=0}^2\Lambda_{1l'}(E_{L-l'}-E_{L+1-l'})\nonumber\\
      &+\Lambda'_{11,00}(E_{L-2}+E_{L}-2 E_{L-1})+
      \Lambda^{\prime L}_{11,0}(E_{L-1}-E_{L})\nonumber\\
      &+\Lambda^{\prime L}_{11,1}(E_{L}-E_{L+1})+
      \Lambda^{\prime R}_{11,0}(E_{L-1}-E_{L}).
\end{align}
These two equations can be rewritten in the general form of
(\ref{49}), provided one adds the boundary conditions
corresponding to (\ref{48c}) and (\ref{48d}). These are in fact
definitions of $E_{-1}$ and $E_{L+2}$:
\begin{align}\label{49d}
\sum_{l'=0}^2\Lambda_{1l'}(E_{1-l'}-E_{2-l'})=&
        \Lambda_{11,10}(E_0+E_2-2 E_1)+\Lambda^L_{11,1}(E_1-E_2)\nonumber\\
      &+\Lambda_{11,00}(E_1+E_3-2 E_2)+\Lambda^L_{11,0}(E_2-E_3)\nonumber\\
      &+\Lambda^R_{11,0}(E_2-E_3),
\end{align}
and
\begin{align}\label{49e}
\sum_{l'=0}^2\Lambda_{2l'}(E_{L+1-l'}-E_{L+2-l'})=&
       \Lambda'_{11,00}(E_{L-2}+E_{L}-2 E_{L-1})\nonumber\\
       &+\Lambda^{\prime L}_{11,0}(E_{L-1}-E_{L})\nonumber\\
      &+\Lambda^{\prime L}_{11,1}(E_{L}-E_{L+1})+
      \Lambda^{\prime R}_{11,0}(E_{L-1}-E_{L}).
\end{align}
Equations (\ref{49}), (\ref{49d}), and (\ref{49e}) can be solved
using the general method of the previous section.

Now consider a special case
\begin{equation}\label{54}
  \lambda_7^j=\lambda_6^4=0.
\end{equation}
The conditions (\ref{51}), and the nonnegativity of the rates,
then lead to
\begin{align}\label{101}
&\lambda_1^2=\lambda_1^4=\lambda_1^6=0,\nonumber\\
&\lambda_2^1=\lambda_2^4=0,\nonumber\\
&\lambda_3^1=\lambda_3^2=\lambda_3^4=\lambda_3^5
=\lambda_3^6=0,\nonumber\\
&\lambda_4^1=\lambda_4^2=\lambda_4^3=0,\nonumber\\
&\lambda_5^1=\lambda_5^2=\lambda_5^3=
\lambda_5^4=\lambda_5^6=0,\nonumber\\
&\lambda_6^1=\lambda_6^2=\lambda_6^3=\lambda_6^4=0,\nonumber\\
&\lambda_7^1=\lambda_7^2=\lambda_7^3=\lambda_7^4=
\lambda_7^5=\lambda_7^6=0,
\end{align}
and
\begin{align}\label{102}
&\lambda_2^5=\lambda_6^5,\nonumber\\
&\lambda_3^7=\lambda_2^5+\lambda_2^6+\lambda_2^7,\nonumber\\
&\lambda_6^7=\lambda_2^7+\lambda_2^3.
\end{align}
Eq. (\ref{49}) then reduces to
\begin{equation}\label{55}
  \dot E_n=AE_{n+2}+BE_{n+1}-(A+B)E_{n},\qquad 1<n<L
\end{equation}
where
\begin{align}\label{56}
  A&:=\lambda_1^5+\lambda_1^7+\lambda_4^5+\lambda_4^7,\nonumber\\
  B&:=\lambda_1^3+\lambda_3^7+\lambda_4^6+\lambda_6^5+\lambda_6^7.
\end{align}
Eq. (\ref{49b}) becomes
\begin{equation}\label{103}
\dot E_1=A'E_3+B'E_2-(A'+B')E_1,
\end{equation}
where
\begin{align}\label{104}
A'&:=\lambda_1^3+2\lambda_1^5+2\lambda_1^7+\lambda_4^5+
\lambda_4^6+2\lambda_4^7-\lambda_5^7,\nonumber\\
B'&:=-\lambda_1^3-\lambda_1^5-2\lambda_1^7+\lambda_3^7-
\lambda_4^5-\lambda_4^6-2\lambda_4^7+2\lambda_5^7+\lambda_6^7,
\end{align}
and eq. (\ref{49c}) becomes
\begin{equation}\label{105}
\dot E_{L}=B''(E_{L+1}-E_{L}),
\end{equation}
where
\begin{equation}\label{106}
B'':=\lambda_1^3+\lambda_1^5+\lambda_1^7+\lambda_2^3+\lambda_3^7
+\lambda_4^5+\lambda_4^6+\lambda_4^7.
\end{equation}
This is in fact a degenerate example of the general case
considered in the previous section. Note that $E_n=0,\quad 1\leq
E_{L+1}$ is obviously a solution. This is expected, since the full
lattice does not evolve, as $\lambda_7^j=0$. Noting that
$E_{L+1}=0$, one can solve (\ref{105}) to obtain $E_{L}$. This is
found to be
\begin{equation}\label{107}
E_{L}(t)=\alpha_{L}e^{-B'' t}.
\end{equation}
Using this, one can solve the equation for $E_{L-1}$, to see that
it contains two exponentials, $\exp(-B'' t)$ and $\exp[-(A+B)t]$.
This is provided $B''\ne A+B$. (Note that in general $B''\leq
A+B$. Equality holds iff $\lambda_2^5=\lambda_2^7=0$.) Let us
assume $B''\leq A+B$ and proceed. It is not difficult to see that
in other $E_n$'s there are also terms like $t^l\exp[-(A+B)t]$. One
can write
\begin{equation}\label{108}
E_n(t)=\alpha_n e^{-B''t}+\sum_{l=0}^{L-n-1}\beta_{n,l}t^l
e^{-(A+B)t},\qquad 1<n\leq L+1
\end{equation}
where
\begin{equation}\label{109}
\alpha_{L+1}=\beta_{L+1,l}=0.
\end{equation}
Putting this in (\ref{55}), one arrives at
\begin{equation}\label{110}
A\alpha_{n+2}+B\alpha_{n+1}+(B''-A-B)\alpha_n=0,
\end{equation}
and
\begin{equation}\label{111}
(l+1)\beta_{n,l+1}=A\beta_{n+2,l}+B\beta_{n+1,l}.
\end{equation}
The solution to (\ref{110}) is
\begin{equation}\label{112}
\alpha_n=\alpha_{L}{{\xi_1^{L+1-n}-\xi_2^{L+1-n}}\over{\xi_1-\xi_2}},
\end{equation}
where $\xi_i$'s are the roots of the equation
\begin{equation}\label{113}
(A+B-B'')\xi^2-B\xi-A=0,
\end{equation}
and $\alpha_{L}$ is arbitrary. The solution to (\ref{111}) is
\begin{equation}\label{114}
\beta_{n,l}=\sum_{s=0}^{l}{{B^{l-s}}\over{(l-s)!}}
{{A^s}\over{s!}}\gamma_{n+l+s},
\end{equation}
where $\gamma_m$'s are arbitrary constants if $1<m<L$, and zero
otherwise.

So far, all $E_n$'s except $E_1$ have been obtained. Using
(\ref{103}), one can also obtain $E_1$. It is seen that $E_1$
contains similar terms and a new exponential term
$\exp[-(A'+B')t]$. So, in general there are only three time
constants in the system, (as long as only the empty intervals are
concerned). It may occur that two of these time constants, or all
of them, are equal. This does not change the general behavior of
the system. Only the degree of the polynomials multiplied in the
exponentials are changed, and the corresponding coefficients can
be calculated similarly.

\vskip\baselineskip

\noindent {\bf Acknowledgement} \\M. Alimohammadi would like to
thank the research council of the University of Tehran, for
partial financial support.

\end{document}